\documentclass[fleqn,usenatbib]{mnras}

\usepackage[T1]{fontenc}

\usepackage{float}

\DeclareRobustCommand{\VAN}[3]{#2}
\let\VANthebibliography\thebibliography
\def\thebibliography{\DeclareRobustCommand{\VAN}[3]{##3}\VANthebibliography}

\usepackage{graphicx}	
\usepackage{amsmath}	
\usepackage{amssymb}	
\usepackage{color}
\usepackage{newtxtext,newtxmath}

\newcommand{\Karmma}{\texttt{KaRMMa}}
\newcommand{\karmma}{\texttt{KaRMMa}}

\newcommand{\Nside}{N_{\rm side}}
\newcommand{\Npix}{N_{\rm pix}}

\usepackage{xcolor}

\newcommand{\avg}[1]{\langle #1 \rangle}

\title[KarMMa]{\Karmma\ - Kappa Reconstruction for Mass Mapping}

\author[Fiedorowicz et al.]{
Pier Fiedorowicz$^{1}$\thanks{E-mail: pierfied@email.arizona.edu},
Eduardo Rozo$^{1}$,
Supranta S. Boruah$^{2}$,
Chihway Chang$^{3,4}$,
Marco Gatti$^{5}$
\\
$^{1}$Department of Physics, University of Arizona, Tucson, AZ 85721, USA\\
$^{2}$Steward Observatory, University of Arizona, Tucson, AZ 85719, USA\\
$^{3}$Kavli Institute for Cosmological Physics, University of Chicago,Chicago, IL 60637, USA\\
$^{4}$Department  of  Astronomy  and  Astrophysics,  University  of Chicago, Chicago, IL 60637, USA\\
$^{5}$Department of Physics and Astronomy, University of Pennsylvania, Philadelphia, PA 19104, USA
}

\date{Accepted XXX. Received YYY; in original form ZZZ}

\pubyear{2021}

\begin{document}
\label{firstpage}
\pagerange{\pageref{firstpage}--\pageref{lastpage}}
\maketitle

\begin{abstract}
We present \Karmma, a novel method for performing mass map reconstruction from weak-lensing surveys. We employ a fully Bayesian approach with a physically motivated lognormal prior to sample from the posterior distribution of convergence maps. We test \Karmma\ on a suite of dark matter N-body simulations with simulated DES Y1-like shear observations. We show that \Karmma\ outperforms the basic Kaiser--Squires mass map reconstruction in two key ways: 1) our best map point estimate has lower residuals compared to Kaiser--Squires; and 2) unlike the Kaiser--Squires reconstruction, the posterior distribution of \Karmma\ maps are nearly unbiased in all summary statistics we considered, namely: one-point and two-point functions, and  peak/void counts.  In particular, \Karmma\ successfully captures the non-Gaussian nature of the distribution of $\kappa$ values in the simulated maps. We further demonstrate that the \Karmma\ posteriors correctly characterize the uncertainty in all summary statistics we considered.
\end{abstract}




\section{Introduction}
\label{sec:intro}

Modern photometric galaxy surveys enable weak gravitational lensing measurements capable of constraining cosmological models with high precision \citep{troxel_etal18, Abbott_2018, hikage_etal19, Hamana_2020, Heymans_2021}. These surveys measure the shapes of tens or even hundreds of millions of galaxies to create weak lensing shear maps of the Universe. As weak lensing is sensitive to all forms of matter along the line of sight, these shear maps probe the mass distribution of the Universe. A related quantity is the weak lensing convergence map (or weak lensing mass map) which measures the matter density field weighted by a lensing kernel and integrated along the line of sight.

Nonlinear structure growth in the evolution of the Universe causes the density field (and thus weak lensing fields) to become non-Gaussian. Because working with mass maps is often simpler than working with shear maps, many methods for extracting non-Gaussian information from the weak lensing signal rely on convergence maps. A non-exhaustive list of these methods includes one-point functions \citep{liuetal19,thieleetal20}, three-point statistics \citep{takada_bispectrum, fu_bispectrum, jung2021integrated}, higher-order moments \citep{Peel_2017, Gatti_2020}, Minkowski functionals \citep{Kratochvil_2012, Petri_2013, Vicinanza_2019}, peak statistics \citep{Dietrich_2010, Kratochvil_2010, Peel_2017, Shan_2017}, and various machine learning methods \citep{Gupta_2018, Fluri_2018, Ribli_2019, Jeffrey_2020_inference}. The extraction of these non-Gaussian signals is of critical importance as they contain significant cosmological information that is highly complementary to that contained in two-point statistics.  For instance, \cite{Gatti_2020} demonstrated that the higher-order moments of the weak lensing maps from a DES-like survey significantly improve the cosmological constraining power of the DES data relative to a standard two-point analysis.

On account of the scientific value of these maps, mass map reconstruction is standard practice for weak lensing surveys \citep{Oguri_2017, Chang_2018}. Traditionally, reconstruction of the convergence field from the shear field is performed by inverting the theoretical relationship between the two fields as initially developed in \cite{KaiserSquires}. Hereafter, we will refer to this reconstruction method as the Kaiser--Squires reconstruction.

While the Kaiser--Squires algorithm is easily the most commonly used, it has two primary drawbacks: the method fails to properly account for the impact of noise and survey masks on the shear fields. Specifically, noisy shear observations may result in apparent mass map fluctuations inconsistent with a cosmological origin. Secondly, as the Kaiser--Squires transformation is non-local, it is necessary to know the shear field on the entire sky. However, because shear maps are limited to a survey window, the unknown shear field outside of the survey mask introduces masking effects in the reconstruction. These masking effects present themselves as additional noise in the reconstructed maps that must be properly attenuated to avoid biases mass map reconstructions near the survey boundaries.

Due to the limitations in the quality of the Kaiser--Squires mass map reconstruction, various alternative methods have been proposed. Many of these techniques involve forward modeling the shear field from the convergence field via the Kaiser--Squires transformation while introducing a Bayesian prior over the convergence field \citep[e.g.][]{alsingetal16,alsingetal17}. Some of these proposed techniques include Wiener filtering \citep{Jeffrey_2018}, sparsity priors \citep{Leonard_2014, Price_2019, Jeffrey_2018}, null B-mode priors \citep{y3_mass_map}, and others \citep{Pires_2020, starck2021weak}. More advanced Bayesian methods aim to forward model the shear field by modeling the initial density field, which is then non-linearly evolved and integrated along the line of sight \citep{Jasche_2013, Porqueres_2021}. Lastly, machine learning-based methods have also been demonstrated to effectively recover the convergence field \citep{shirasaki2019decoding, Jeffrey_2020, hong2021weaklensing}.

While we believe that the most physically correct approach to reconstructing the convergence field relies on simulation-based forward modeling \citep[as advocated for instance in][]{Porqueres_2021}, there is still significant value in the development of fast, approximate reconstruction schemes. Specifically, approximate reconstruction methods can be used to study how to incorporate systematics in the forward modeling approach within a much simplified and significantly more numerically efficient framework. Moreover, it is not yet clear to what degree the use of simplified numerical models will compromise our ability to reconstruct accurate mass maps.  If the biases incurred due to the use of analytic approximations are small compared to statistical uncertainties, then the tremendous gain in speed of analytic methods would make them extremely attractive.

Here, we propose to replace the simulation-based model of the convergence field with a fast, analytical approximation.  The simplest such approximation models the convergence field as a Gaussian random field. In this case, a Wiener filter of the Kaiser--Squires reconstruction produces the maximum a posteriori estimate for the field. This is computationally expeditious, but the reconstruction fails to correctly recover the non-Gaussian fluctuations due to non-linear evolution \citep{Jeffrey_2018}. Motivated by the fact that a lognormal approximation results in a much more accurate description of the convergence field \citep{clerkin_etal17}, we have opted for forward modeling the latter as a homogeneous and isotropic lognormal random field. This is similar to the approach in \cite{B_hm_2017}, though we note there the prior applies to the 3D density field, whereas we use the prior to describe the 2D convergence field \citep[for other applications of the lognormal prior see e.g.][]{jaschekitaura10,kitauraetal10}.

In this paper, we introduce \Karmma\ (Kappa Reconstruction for Mass Mapping), a new method for performing mass map reconstruction. \Karmma\ is similar to some of the previously mentioned techniques in that it is a Bayesian reconstruction method. \Karmma\ introduces a physically motivated lognormal prior \citep{coles_jones_91} on the convergence maps. As a result, \Karmma\ significantly improves reconstruction quality over Kaiser--Squires, correctly captures the non-Gaussianities of the convergence maps, and recovers the correct two-point statistics. Additionally, unlike many similar techniques, \Karmma\ generates sample convergence maps from the posterior distribution rather than a single "best fit" map. That is, \Karmma\ fully quantifies the uncertainties in our posteriors.  We caution that the current \Karmma\ algorithm explicitly assumes a cosmological model when implementing the lognormal prior.  Consequently, \it the current \Karmma\ maps cannot be used for cosmological inference. \rm  In future work, we intend to enable joint sampling of cosmology and mass maps.


\section{Model}
\label{sec:model}

We forward model the observed shear field as a noisy realization of an underlying, noiseless convergence field. The latter is obtained as the ``$\kappa$ to $\gamma$'' (or forward) Kaiser--Squires transformation of the true convergence field.  Conceptually, our model parameters are the values of the convergence field in pixels in the sky, which are modeled as a realization of a lognormal random field.  In practice, our parameterization is slightly more complicated for reasons that will be made apparent momentarily.  For now, it is best to start with this ``conceptual'' parameterization.  Specifically, the convergence field $\kappa$ is assumed to be a non-linear transformation of a Gaussian random field $y$ such that \citep[][]{hilbertetal11,Xavier_2016}
\begin{equation}
y_i \equiv \ln(\kappa_i + \lambda) .
\label{eq:y}
\end{equation}
where $y_i$ is the value of $y$-field in pixel $i$.  The parameter $\lambda$ above is the "shift" parameter  and can be interpreted as the minimum possible value that $\kappa$ can take in a pixel.  By definition, the distribution for $y$ is
\begin{equation}
    P(\vec{y}) \propto \exp \left[ -\frac{1}{2} (\vec{y} - \vec{\mu})^\top \mathbf{\Sigma}^{-1} (\vec{y} - \vec{\mu}) \right].
\label{eq:Py}
\end{equation}
The covariance matrix between pixels, $\mathbf{\Sigma}$, is fully specified by the correlation function of the \it convergence \rm field $\xi(\theta)$ and the shift parameter $\lambda$ via,
\begin{equation}
\label{eq:covariance}
    \Sigma_{ij} = \ln \left( \frac{\xi(\theta_{ij})}{\lambda^2} + 1\right).
\end{equation}
Lastly, the parameter $\mu$, is constrained by the fact that the mean convergence is zero.  Specifically,
\begin{equation}
    0 = \left< \kappa_i \right> = \exp \left( \mu_i + \frac{\Sigma_{ii}}{2} \right) - \lambda ,
\label{eq:ymean}
\end{equation}
and therefore
\begin{equation}
    \mu_i = \ln(\lambda) - \frac{\Sigma_{ii}}{2}.
\label{eq:yvar}
\end{equation}

Together, equations~\ref{eq:y} through \ref{eq:yvar}, specify the probability distribution of the Gaussian random field $y$ in terms of the shift parameters $\lambda$ and the convergence power spectrum $C_{\ell}$.  The latter can be readily computed using publicly available tools such as CAMB \citep{Lewis_2000} or CLASS \citep{Blas_2011}. To determine the lognormal shift parameter, we fit for the value of $\lambda$ that minimizes the mean squared error between the lognormal pdf and the empirical pdf from the simulations described in section~\ref{sec:sim_tests}. The convergence field of interest is obtained as the inverse of the non-linear transformation in equation~\ref{eq:y}, where $y$ is subject to the Gaussian prior of equation~\ref{eq:Py}.

While we aim to reconstruct the pixelized convergence field, our observable is a pixelized shear field. The shear field $\gamma$ generated by the convergence field $\kappa$ is calculated using the usual $\kappa$ to $\gamma$ Kaiser--Squires transformation in harmonic space,
\begin{equation}
    \gamma_{\ell,m} = - \sqrt{\frac{(\ell + 2) (\ell - 1)}{\ell (\ell + 1)}} \kappa_{\ell,m}.
\end{equation}
The observed shear field is then modelled as a noisy realization of this predicted shear field. The likelihood distribution is therefore a Gaussian distribution where the shape noise in pixel $i$ has a variance $\sigma_i^2 = \sigma_\epsilon^2 / N_i$ where $\sigma_\epsilon$ is the shape noise per source and $N_i$ is the number of source galaxies in pixel $i$.  We assume the noise is uncorrelated across pixels.

The full posterior distribution for our model is
\begin{equation}
    P(\vec{y} \mid \vec{\gamma}_\mathrm{obs}) \propto P(\vec{y}) \times \exp \left [ -\frac{1}{2} \sum_i \frac{
    (\gamma_{i,\mathrm{obs}} - \gamma_i(\vec{\kappa}))^2}{\sigma_i^2} \right].
\end{equation}

We note that for a map with $\Npix$ pixels, the covariance matrix $\Sigma$ characterizing the prior is $\Npix\times \Npix$.  For this reason, maps with more than $\sim 10^5$ pixels become numerically intractable.  For the current study, we have limited ourselves to maps of $\approx 30'$ resolution, corresponding to HEALpix $\Nside=128$. At this resolution, our maps contain approximately $17,000$ pixels in total, including the buffer regions around the survey edges as described in section~\ref{sec:mask_effects}.  Finally, for reasons that will be made apparent in section~\ref{sec:res_err}, our reconstructed maps are band limited to modes with $\ell \leq 2~\Nside$.

\begin{figure*}
    \centering
	\includegraphics[width=\textwidth]{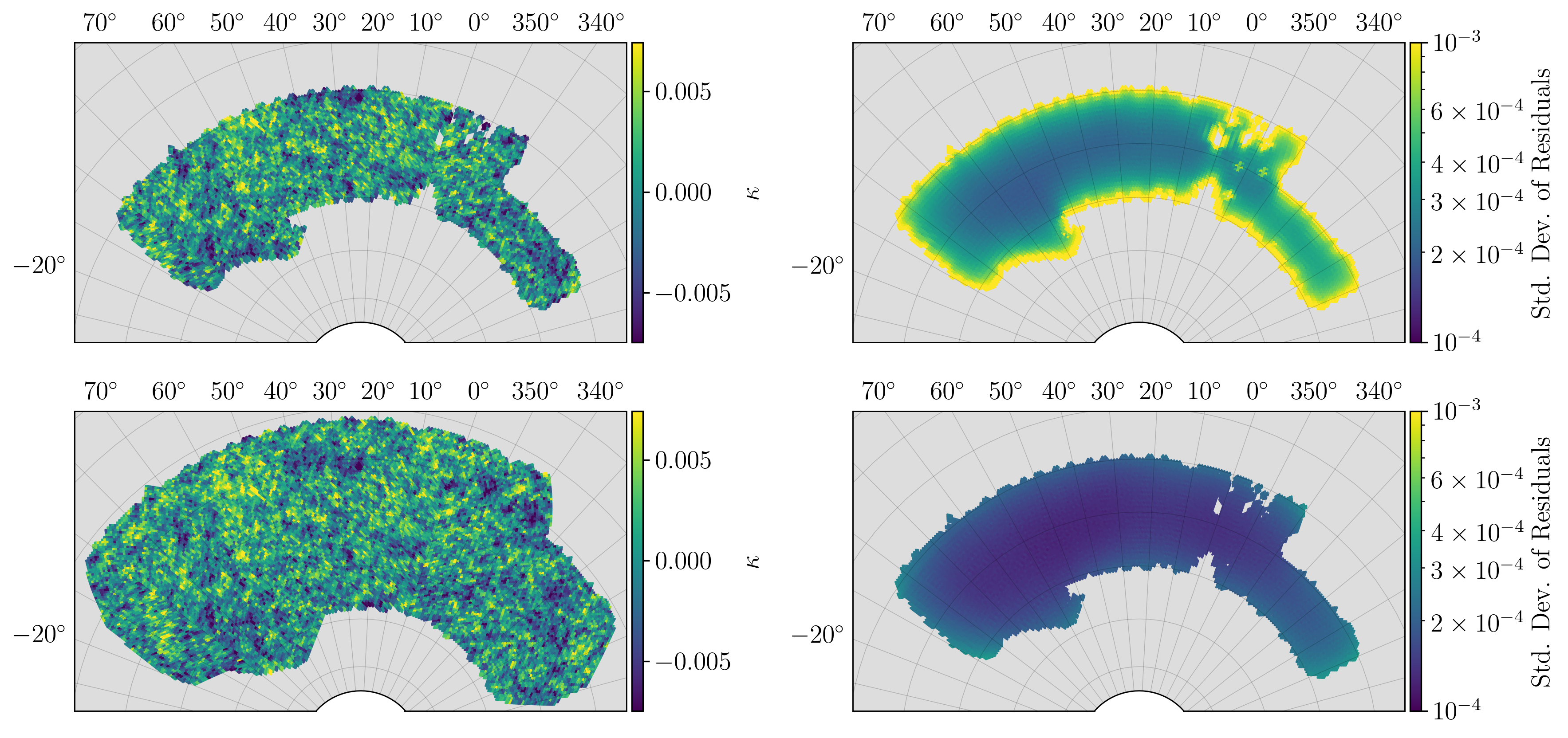}
    \caption{Illustration of masking effects and how \Karmma\ suppresses them. Top Left: An example $\kappa$ map with the DES Y1 mask applied. When the convergence field is set to zero outside the mask, the shear field produced by the input convergence field is modified relative to that obtained using the full-sky convergence field.  We refer to this difference as ``masking noise.''  Top Right: Masking noise in the shear field using convergence fields restricted to the survey mask, as estimated using 1000 mocks. Bottom Left: Sample $\kappa$ map with the DES Y1 mask and an additional 10 pixel buffer applied. Bottom Right: Masking noise recovered after adopting the proposed buffer region. The color scale is the same between the maps.}
    \label{fig:masking_effects}
\end{figure*}

In practice, we reparameterize the convergence map by diagonalizing the prior covariance matrix $\mathbf{\Sigma}$ using Singular Value Decomposition (SVD). Our model parameters are therefore the coefficients of the basis vectors spanning the space of convergence maps.  This has two benefits: 1) While not strictly necessary, working in a diagonal basis makes sampling simpler; and 2) SVD allows us to truncate singular modes to ensure numerical stability in the posterior calculation. Because the resulting basis vectors are orthonormal, one can think of these basis vectors as being similar to the $a_{lm}$ coefficients of the $y$-field.  Truncation of singular modes is performed by plotting the singular values (normalized by the largest singular value) of the covariance matrix (not shown). We find a distinct several order of magnitude drop off in the singular values after the first $\approx 6000$ modes. Modes beyond this point do not contribute meaningfully to the map statistics and are not included in the model. This agrees well with the expected number of principal components. \footnote{Specifically, HEALpix $\Nside=128$ maps can resolve up to $l_{\rm max}=256$ mode, and can therefore be fully specified by the value of $2l_{\rm max}(l_{\rm max}+1)\approx 1.3\times 10^5$ real coefficients.  The mass maps in this work are $\approx 1,500\ {\rm deg}^2$, so we'd expect the number of coefficients required to characterize our mass maps is $\approx (1.3\times 10^5)\times (1,500/40,000)\approx 5,000$.}


\section{Numerical Systematics}
\label{sec:num_sys}

We discuss three sources of numerical systematics in our reconstruction algorithm.  Of these, the first two must be adequately treated to avoid biases in the posterior distributions, whereas the last one has only a minor impact on the posteriors.

\subsection{Masking Effects}
\label{sec:mask_effects}
The Kaiser--Squires transformation relating the convergence and shear fields is most easily performed in harmonic space.  The harmonic decomposition of a map can only be performed on the entire sphere, so one usually ``zeroes out'' all pixels outside the survey boundary for the purposes of computing the harmonic transforms.  This cut introduces artifacts near the edges of the mask in the recovered map \citep[e.g.][]{Chang_2018}.

Because we forward model the convergence field, we can readily include as parameters the values of the convergence field outside of the survey footprint, thereby significantly reducing masking effects. For instance, \cite{Mawdsley_2020} applied this technique by modeling the full sky convergence field. However, a full sky reconstruction is computationally demanding and entirely unfeasible in our code, even at modest resolutions.  Therefore, we strike a balance between computational feasibility and minimizing masking effects by adding a buffer region around the mask within which we perform our reconstruction. Including this buffer region ensures that the induced effects are primarily limited to the edges of the buffered mask, rather than the survey mask (see figure \ref{fig:masking_effects}).\footnote{All map visualizations in this paper are created using the publicly available SKYMAPPER code: \url{https://github.com/pmelchior/skymapper}} The size of this buffer region is selected such that increasing the size of the buffer provides no appreciable reduction in masking artifacts. The remaining noise due to masking effects is also accounted for in our reconstruction, as detailed in Appendix \ref{sec:bias}.

\subsection{Resolution Error}
\label{sec:res_err}

The resolution of HEALPix maps is set by the variable $\Nside$. A HEALPix map of $\Nside = 128$ roughly corresponds to pixels of $30'$ resolution. Doubling the $\Nside$ will result in a map where each pixel is subdivided into four pixels of equal area. When performing spherical harmonic transformations with HEALPix, modes with $\ell > 2~\Nside$ are poorly resolved. Consequently, spherical transforms can ``leak'' power from unresolved modes into larger scales.  This problem is usually referred to as aliasing.  We therefore limit our map reconstruction to well-resolved modes with $\ell \le 2~\Nside$, resulting in a band-limited map. However, since the observed shear map has no such band-limit, the likelihood in our posterior is incorrect. We calibrate the uncertainty due to lost power using lognormal mocks generated with Flask \citep{Xavier_2016}. This uncertainty is added as an additional source of theoretical noise to the likelihood term, and accounts for the uncertainty associated with the missing power. The variance of this additional noise term (including masking effects and missing power) is 10-15\% of that for the shape noise term. As such, ignoring this term would introduce non-negligible bias in the reconstruction. The details of this calibration are important for mitigating bias in our posterior maps, so we discuss these at some length in Appendix \ref{sec:bias}.

\subsection{Pixel Geometry}

We use the HEALPix \citep{Gorski_2005} pixelization scheme to pixelize the curved sky. The primary benefit of the HEALPix scheme is that all pixels have equal areas. However, this comes at the cost of each pixel having a different shape. Naively, one might expect the covariance matrix between two pixels in equation~\ref{eq:covariance} would depend only on the separation between the pixels.  The fact that different pixels have different shapes implies that this is not quite true, and one must account for pixel geometry.  To compute $\xi_{ij}$, we subdivide the pixels $i$ and $j$ into sets of pixels at a higher resolution, $i'$ and $j'$. We then compute the covariance between all subpixels $i'$ and $j'$ according to equation~\ref{eq:covariance}, that is, assuming the covariance matrix between these high-resolution pixels depends only on separation. Because the value of the convergence in pixel $i$ is the average value across all subpixels, $\kappa_i = \frac{1}{N} \sum_{i'} \kappa_{i'}$, we can compute the covariance between the original coarse pixels via
\begin{equation}
    \mathrm{Cov}(\kappa_i, \kappa_j) = \frac{1}{N^2} \sum_{i'}\sum_{j'} \mathrm{Cov}(\kappa_{i'}, \kappa_{j'}).
\end{equation}
By increasing the subdivision depth to higher resolutions, one can account for the pixel geometry in the prior covariance matrix to arbitrary accuracy.  For an $\Nside=128$ map, the geometry-corrected variance computed at $\Nside=1024$ results in a $0.57\%$ median absolute deviation and $3.0\%$ maximum deviation from the naive variance over the DES Y1 mask. In practice, we do not find that this term significantly impacts the analysis; however, we retain it for correctness.


\section{Posterior Sampling}

Our parameterization of the convergence maps can have $\sim 10^5$ or even more parameters.  Consequently, efficient sampling of this high-dimensional parameter space is strictly necessary. We achieve efficient sampling using Hamiltonian Monte Carlo (HMC) sampling \citep{neal2012mcmc}.

HMC treats the posterior probability distribution as a potential energy function, and introduces momenta as nuisance parameters. The HMC sampler evolves the state of the system according to Hamiltonian mechanics to reach a new proposal sample state. As this process conserves the total ``energy,'' or in our case, probability, the sampler accepts the proposed state with a probability of unity and discards the nuisance parameters. In practice, the acceptance rate is slightly below unity due to numerical integration errors. 

The efficiency of HMC samplers is critically sensitive to the choice of the mass matrix $\mathbf{M}$ used in the kinetic energy term of the Hamiltonian
\begin{equation}
    T = \frac{1}{2} \vec{p}^\top \mathbf{M}^{-1} \vec{p}.
\end{equation}
Failure to choose an appropriate mass matrix can result in a highly inefficient sampling of the posterior. In such a case, a limited sampling of the posterior will result in a biased chain that does not cover the whole posterior distribution. Optimal efficiency is typically achieved by setting the inverse mass matrix to the covariance matrix of the target distribution \citep{tayloretal08,neal2012mcmc}. Naturally, the momenta $\vec{p}$ should be randomly drawn from the covariance matrix $\mathbf{M}$. For our model, we set the inverse mass matrix to the diagonalized prior covariance matrix. With this setup, we find our posterior chains have an acceptance rate of $0.65$ and a correlation length of $3.2$ samples. Using 1,000 samples per chain, this corresponds to an effective sample size of $\sim 300$ samples.  While this number of samples is clearly insufficient for a full sampling of the map posterior space, we have used a chain with 100 times more samples to verify that our default chain length is sufficient for accurately recovering all summary statistics from our maps.  We attribute this result to the fact that the sky area we use is quite large, so any individual map already contains many independent realizations of the convergence field within it.

The HMC sampler used in this work is a custom implementation of the standard HMC sampling algorithm using the PyTorch framework for CUDA acceleration and automatic gradient calculations. This allows us to generate independent samples at a rate of $16,500$ samples/hour using a consumer NVIDIA RTX 2070 Super graphics card.


\section{Simulation Tests}
\label{sec:sim_tests}

We test our method using the suite of 108 full-sky dark-matter simulations of \citet{Takahashi_2017}.  These simulations were generated using a fixed flat $\Lambda$CDM cosmology with parameters $\Omega_\Lambda=0.279$, $\Omega_b=0.046$, $h=0.7$, $\sigma_8=0.82$, and $n_s=0.97$. The mocks are provided at a HEALPix resolution of $\Nside=4096$.  Because our method is computationally restricted to a lower resolution of $\Nside=128$, we downgrade the simulated maps by averaging the shear of the high resolution maps within each $\Nside=128$ pixel.

The simulations are given as $\kappa$ maps at several redshift slices. From these maps, we construct DES Y1-like mass maps by adding the individual slices weighted by the Y1 non-tomographic $dn/dz$ \citep{Hoyle_2018}. To generate observed shear maps, we add shape-noise consistent with the DES Y1 shear maps \citep{Zuntz_2018} to the downgraded true shear maps. This noise is constructed by Poisson sampling a galaxy number count for each pixel with a mean of 4.5 galaxies per square arcmin. Gaussian noise is added to $\gamma_i$ for each pixel $i$ by sampling the shape noise from $\sigma_i = \sigma_\gamma / \sqrt{N_i}$ where $\sigma_\gamma = 0.28$ is the standard deviation of galaxy shapes.  In this way, we produce 108 simulated Y1-maps at $\Nside=128$ resolution.  We emphasize that the downgraded shear maps include high-$l$ (i.e. $l> 256)$ power.

We wish to compare our recovered convergence maps to the input mass maps from the numerical simulations. Recall that our model results in band-limited maps at $\Nside=128$ resolution.  Therefore, to compare to simulations we first apply a low-pass filter to the $\Nside=4096$ convergence map such that $a_{\ell,m}=0$ for all modes unresolved in our analysis ($\ell >2\times 128$).  We then downgrade the low-pass filtered $\Nside=4096$ map to $\Nside=128$.  This low-pass filtered downgraded map is the ``truth'' that our algorithm should recover. Note that while our ``truth'' map is low-pass filtered, our synthetic shear maps are not.  Also, we have found that low-pass filtering the high resolution map before downgrading is important to avoid aliasing in the final ``truth'' maps.

In what follows, we will compare our results against the standard Kaiser--Squires algorithm.  Specifically, we analyze each of the 108 simulated DES-Y1 data sets, and compare: 1) the residuals between ``truth'' (i.e. the low-resolution simulated convergence maps) and each of the reconstructions (Kaiser--Squires and \Karmma); and 2) the one- and two- point distributions of the true and reconstructed maps.


\section{Results}
\label{sec:results}

\begin{figure*}
    \centering
	\includegraphics[width=\textwidth]{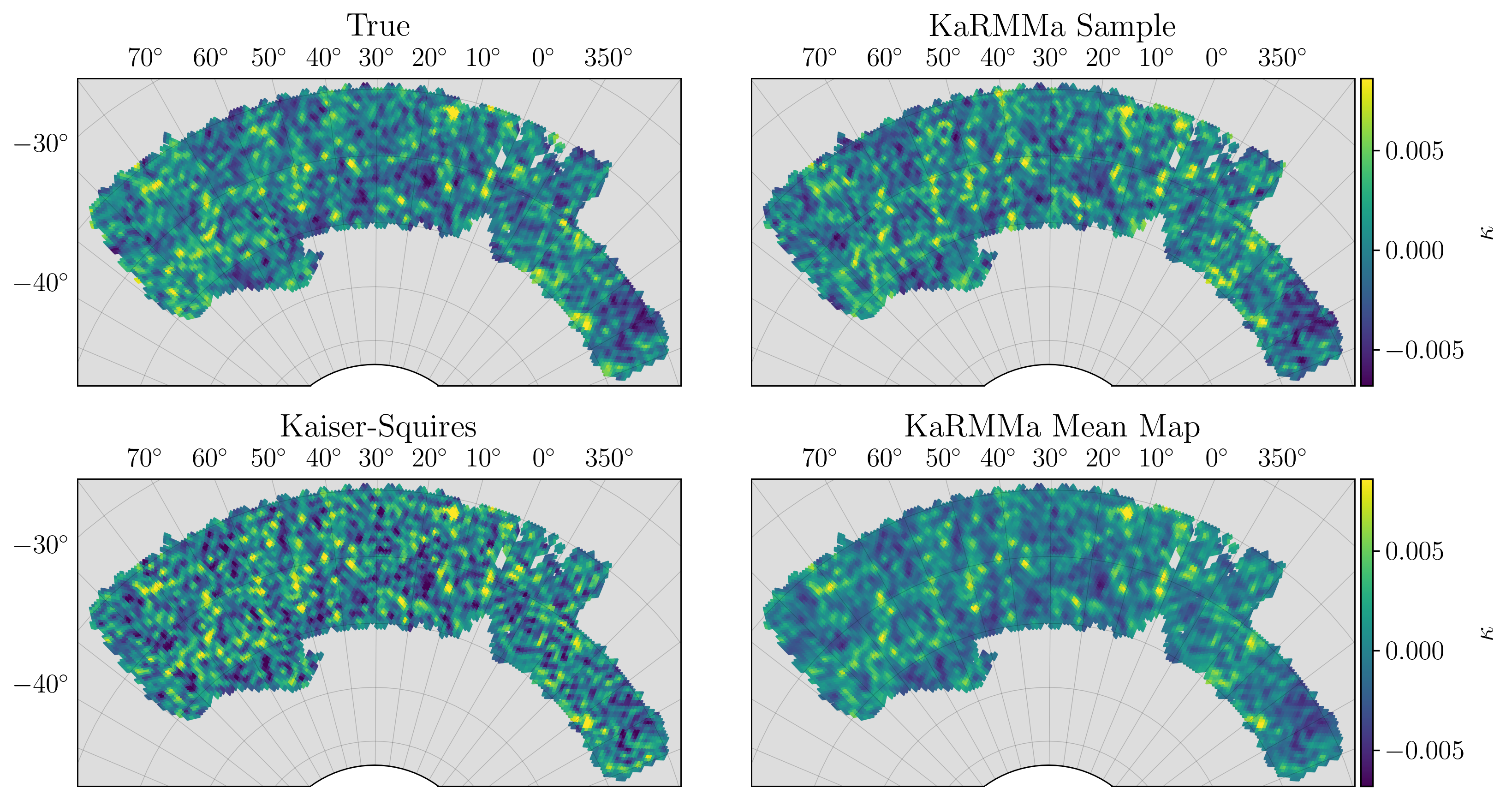}
    \caption{Visual comparison of mass map reconstructions for a single mock realization. Top Left: True mass map from the simulation. Top Right: One posterior sample map from \Karmma. Bottom Right: Mean map from \Karmma. Bottom Left: Kaiser--Squires reconstruction. Note the Kaiser--Squires reconstruction is far noisier than the true mass map, with excess small scale structure. By contrast, the \Karmma\ sample captures structure at the correct locations and physical scales when compared to the true convergence map. Bottom right: the mean of the posterior distribution of \Karmma\ maps.  Note this mean map smooths out prior-dominated scales, thereby suppressing small-scale structure.}
    \label{fig:vis_comp}
\end{figure*}

Before we present our results we need to setup some important conventions.  In all the figures in this section, different colors correspond to different types of maps as follows:
\begin{itemize}
    \item green: statistics for the Kaiser--Squires reconstruction
    \item orange: statistics for the mean map computed from the \Karmma\ posteriors.
    \item purple: statistics for individual maps in the posterior distribution of \Karmma\ maps.
\end{itemize}
The difference between orange and purple is important.  The ``mean map'' (orange) refers to the process of taking all the maps in the \Karmma\ posterior, and averaging them out to obtain a single mean map. This mean map can be thought of as the single best-point estimate derived from our posterior.  In practice, this mean map will not exactly coincide with the true maximum posterior unless the posterior is Gaussian.  By contrast, the ``average statistic'' in individual \Karmma\ samples (purple) corresponds to computing the statistic of interest for each of the \Karmma\ maps in the posterior distribution, and then averaging across all posterior maps.  It is the latter operation that is relevant for testing whether our posterior distributions are biased or not \citep[e.g.][]{hoggetal10}.

Because the \Karmma\ maps are low-passed filtered, we compare our posteriors to the statistics of the low-pass filtered simulation maps (see section~\ref{sec:sim_tests}).  Further, we also low-pass filter the Kaiser--Squires reconstruction.  This way, we can perform a true ``apples-to-apples'' comparison between the two methods.  We note that the low-pass filtering significantly reduces the obvious high-frequency noise that is otherwise present in the Kaiser--Squires reconstruction.

Figure~\ref{fig:vis_comp} provides a visual comparison between the maps generated by Kaiser--Squires/\Karmma\ and truth for a single mock realization. Note that the Kaiser--Squires reconstruction appears to capture the correct locations of large under/over-densities. However, it exhibits a significant amount of noise, particularly at smaller scales. By contrast, the sample map from \Karmma\ not only captures features at the correct locations, the relative amount of power at different scales appears to be qualitatively similar to that of the true map. When comparing the mean map from the \Karmma\ posterior, we find that it similarly recovers the locations of large over/under-densities, but that small scale fluctuations are effectively smoothed out. This map is qualitatively similar to the Wiener-filtered map, where prior-dominated scales are smoothed over. Indeed, as mentioned earlier, we can think of the mean map as the best point-estimate for the convergence field.

\begin{figure}
    \centering
	\includegraphics[width=\columnwidth]{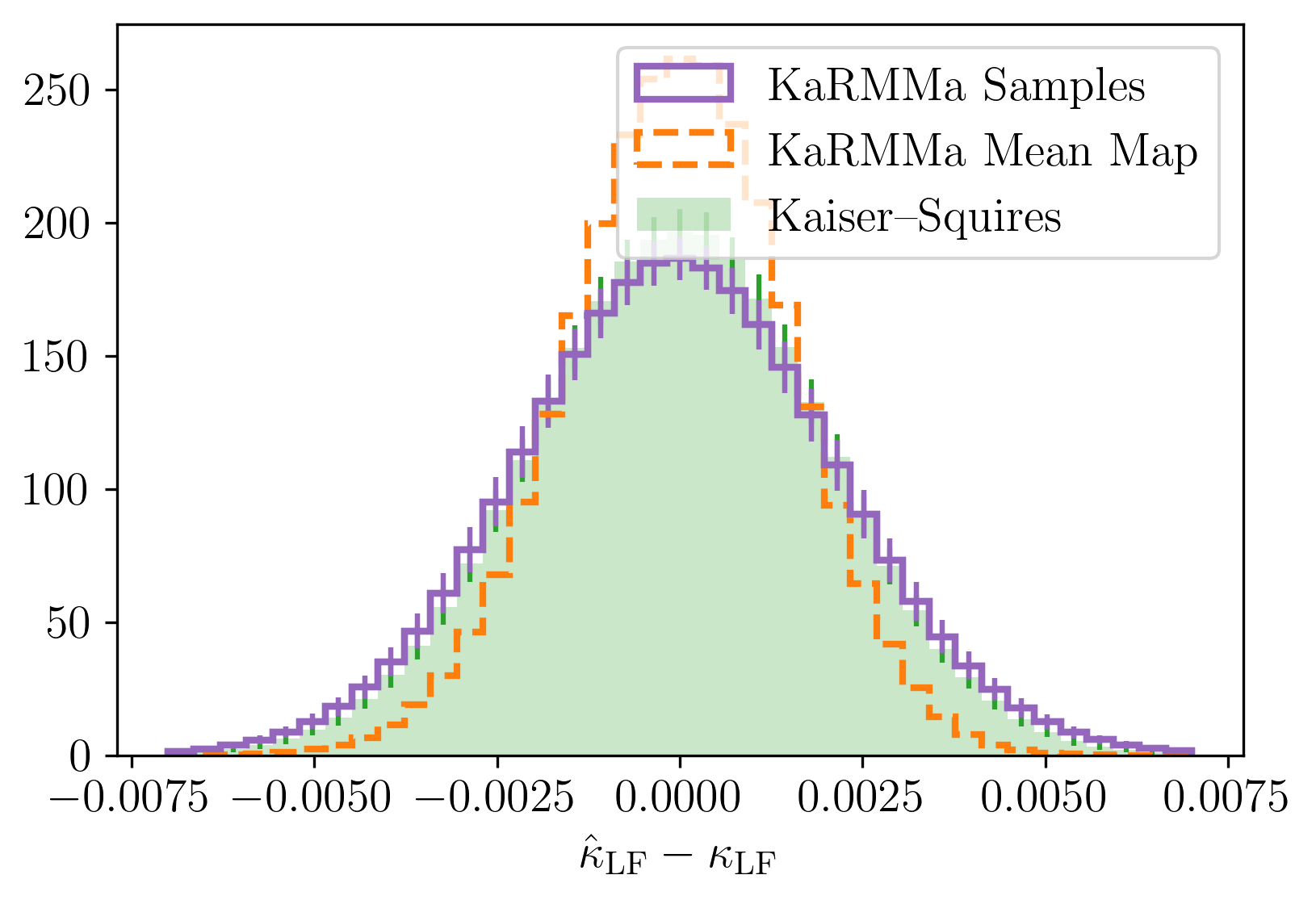}
    \caption{Distribution of pixel-wise residuals between the Kaiser--Squires and \Karmma\ reconstructions. The orange histogram shows the corresponding residuals for the mean of the maps in the posterior, or ``mean map'' for short. The purple map shows the distribution of residuals averaged over maps sampled from the \Karmma\ posterior.  We find average Pearson correlation coefficients between the true mass maps and the reconstructions of 0.77 for Kaiser--Squires, 0.68 for \Karmma\ samples, and 0.82 for \Karmma\ average.}
    \label{fig:resids_hist}
\end{figure}

We now validate the visual impressions from Figure~\ref{fig:vis_comp} at a quantitative level.  Figure \ref{fig:resids_hist} compares the distribution of residuals between the true and reconstructed $\kappa$ maps.  We see from the figure that that the distribution of residuals for the mean \Karmma\ map (orange) is significantly tighter than that of the Kaiser--Squires reconstruction (green).  That is, the mean map is a much better point-estimate of the mass map than the Kaiser--Squires map.  As per the above discussion, this makes sense: the averaging procedure dampens unresolved modes, thereby reducing noise.  By contrast, the distribution of residuals for \Karmma\ sample maps is slightly broader than that of the Kaiser--Squires map. This is surprising, as it demonstrates that a random realization from the \Karmma\ posterior --- which, in accordance to our prior, has added noise relative to our best point estimate ---  is nevertheless nearly as good a point-estimate of the true mass map as the original Kaiser--Squires reconstruction! However, the fact that even by eye the Kaiser--Squires map is qualitatively different from the true convergence map (see Figure~\ref{fig:vis_comp}) implies its noise fluctuations do not have the spatial structure expected from a true convergence map. We will confirm this momentarily.  Before we do so, we quantify the fidelity of the reconstructed maps using the Pearson correlation coefficient. The correlation between the true maps and each of the reconstructions is $0.771 \pm 0.009$ for Kaiser--Squires, $0.823 \pm 0.010$ for the mean \Karmma\ map, and $0.676 \pm 0.014$ for \Karmma\ sample maps.  Error bars represent the error in the mean across all 108 simulated maps. The fact that the correlation coefficient between \Karmma\ samples and the true convergence field is significantly lower than that of the mean map is expected: the prior term in our posterior adds noise to the mean map so as to recover the expected clustering statistics of a cosmological convergence map.  It is this noise which is responsible for decreasing the correlation coefficient relative to our mean mass map.

\subsection{\Karmma\ Posterior Validation}
\label{sec:posterior_validation}

We wish to verify that the posterior distributions returned by \Karmma\ correctly capture  measurement uncertainties.  We consider in particular the uncertainties in the recovered one point function, two point functions, and peak/void counts, as determined from the posterior distribution of convergence maps. To do so, consider first the posterior of the one-point function of the convergence map.  Assuming this posterior is Gaussian, it is completely characterized by its mean and covariance matrix.  We test this hypothesis as follows: 1) given a synthetic data set, for each sample map in the chain we compute its one-point function within the survey footprint.  The one-point function is evaluated using 19 convergence bins, resulting in a binned histogram representation of the one-point function. The number of bins is a compromise between the desire for a large number of bins to adequately test the distribution, and the necessity of having sufficient statistics within each bin.  2) Using the full posterior distribution of maps, we compute the average one-point function in each of these bins and the corresponding covariance matrix. 3) For each of our 108 synthetic data sets, we evaluate the $\chi^2$ statistic for the difference between the one-point function of the true convergence field and the mean one-point function determined from our posterior. The chi-squared statistic for data vector $\vec{d}$ of simulation $i$ can be expressed as
\begin{equation}
    \chi^2_i = \Delta \vec{d}_i^\top \Sigma^{-1}_i \Delta \vec{d}_i
\end{equation}
where $\Delta \vec d_i = \avg{\vec d_i}_{\rm HMC} - \vec d_{i,{\rm true}}$, and $\avg{\vec d_i}_{\rm HMC}$ is the mean of the posterior for the summary statistics $\vec d$ for simulation $i$ computed from our HMC samples. Likewise, the covariance matrix $\Sigma_i$ are estimated from the HMC samples of $\vec d_i$ constructed using the sampled maps. When calculating the precision matrix, we account for the expected Hartlap correction-factor to the $\chi^2$ statistic \citep{Hartlap_2006}. If the posterior for the one-point function were Gaussian, the distribution of $\chi^2$ values across our 108 synthetic observations ought to follow a $\chi^2$ distribution with 19 degrees of freedom.  Using a similar algorithm, we can also test whether the posterior of the two-point function and peak/void counts of the convergence maps are consistent with a Gaussian distribution characterized by the mean and covariance matrix derived from the sample posteriors.  We test the posteriors for the two-point function in both real and harmonic space, using 19 bins for consistency in the number of bins across all summary statistics, which makes visualization easier.

\begin{figure*}
    \centering
    \includegraphics[width=\textwidth]{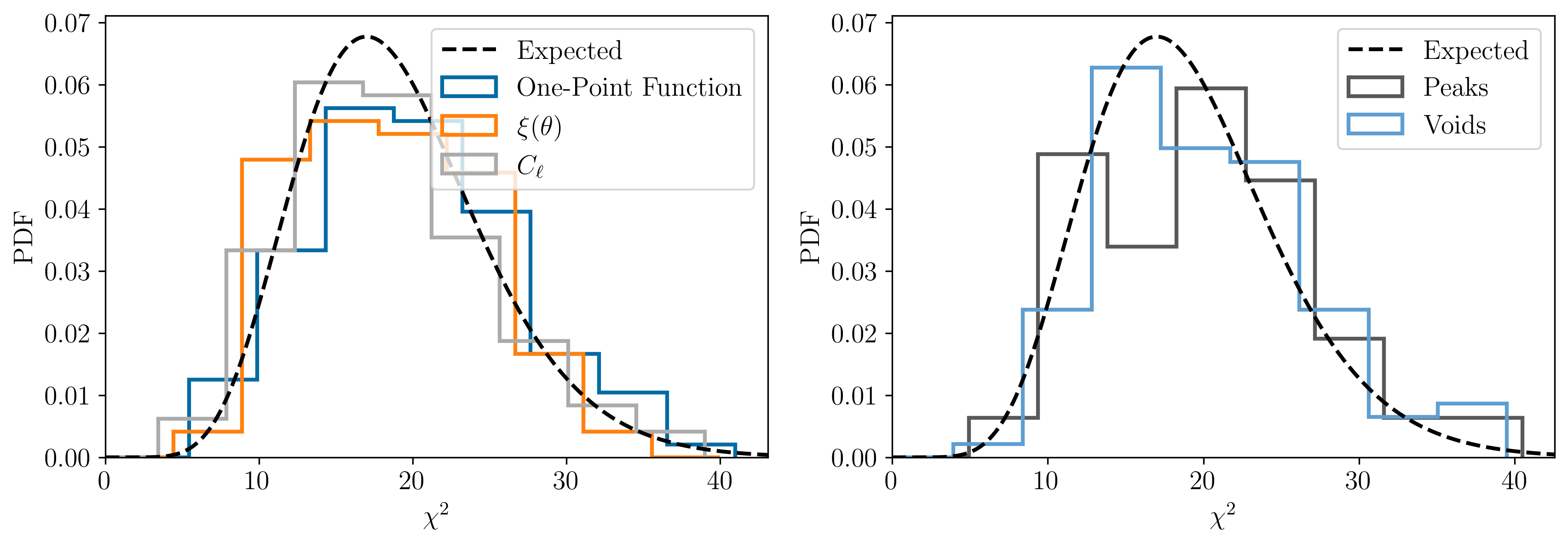}
    \caption{Left: Distribution of the $\chi^2$ statistics for the difference between the mean one-point/two-point statistics from \Karmma\ and the true statistics across the 108 mock realizations. Right: Distribution of $\chi^2$ statistics for peak/void counts. In both panels, the bins for different observables are slightly offset from each other for visualization purposes.  The $\chi^2$ values are computed using covariances estimated directly from the \Karmma\ chains. In black we show the expected $\chi^2$ distribution with 19 degrees of freedom (all statistics were computed in 19 bins). The excellent agreement between the empirical distribution of $\chi^2$ values and the expected $\chi^2$ distribution demonstrates that the \Karmma\ posterior accurately captures the uncertainty in our reconstructions.}
    \label{fig:chisqs}
\end{figure*}

\begin{figure*}
    \centering
	\includegraphics[width=\textwidth]{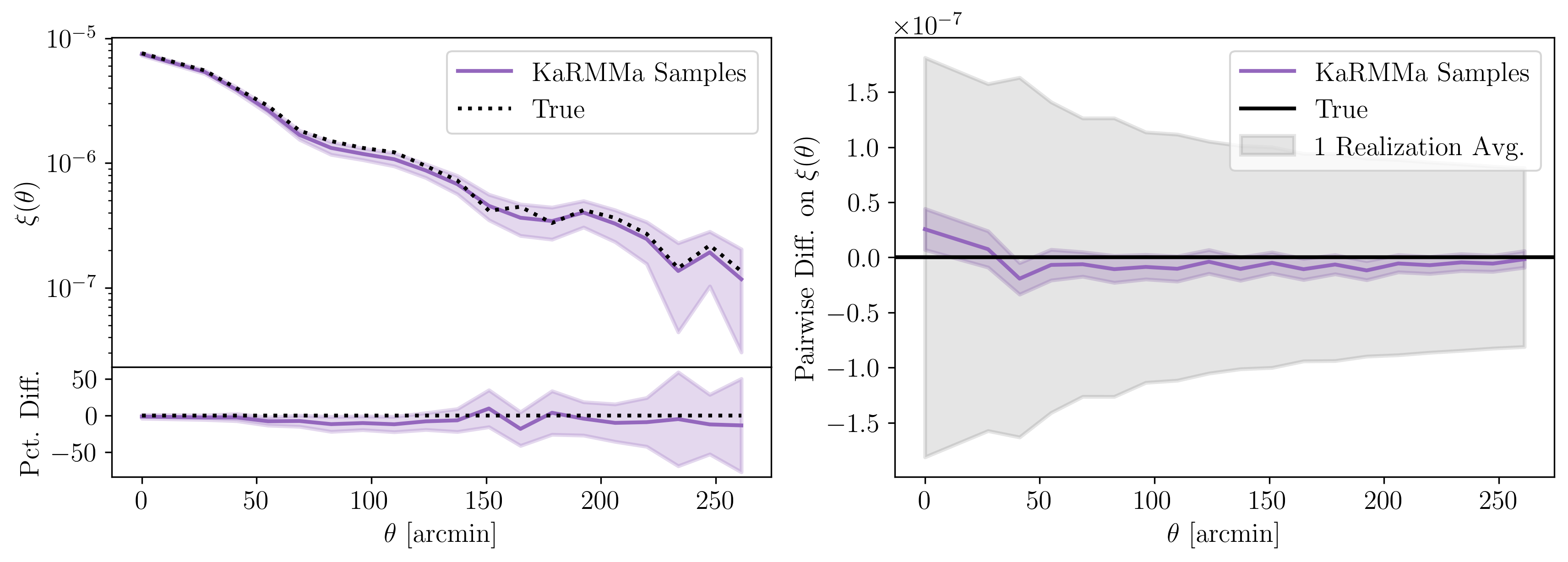}
    \caption{Left: Comparison between the correlation function reconstructed by \Karmma\ (purple) and the true (black) correlation function in a single mock realization. Purple bands represent the $1\sigma$ uncertainty from the \Karmma\ posterior. Note that the true correlation function falls within the posterior uncertainties. Right: The purple line shows the difference between the mean correlation function from the \Karmma\ posterior and the true correlation function, averaged across all 108 synthetic data sets. Purple bands represent the $1\sigma$ error on the mean. The small deviation from zero in the above plot indicates marginal evidence ($2.9\sigma$) of bias in our reconstruction. This bias is negligible compared to observational uncertainties in a single mock (grey band).}
    \label{fig:samp_cf_and_bias}
\end{figure*}

Figure~\ref{fig:chisqs} compares the distribution of chi-squared statistics from \Karmma\ for each summary statistic to that of the expected chi-squared distribution. It is clear from the plot that the \Karmma\ posteriors adequately characterizes measurement uncertainties.  That is, the simulation-to-simulation variance in the summary statistics is accurately captured by the width of the posterior distributions estimated from a single simulation map. Under the Kolmogorov--Smirnov test, we find that the empirical chi-squared distributions are consistent with the expected distribution with $p$-values of $0.11$, $0.62$, and $0.52$ for the one-point function, correlation function, and power spectrum respectively. The $p$-values for the chi-squared distributions for peak and void counts are $0.027$ and $0.021$ respectively.  These corresponds to a $\sim 2\sigma$ detection of a difference between the two distributions, though we caution that there is significant noise in this estimate: in each case, there are only a few hundred pixels spread across the 19 bins. As further illustration of the fact that our posteriors correctly capture measurement uncertainties, in the left panel of figure~\ref{fig:samp_cf_and_bias} we compare the the width of the \Karmma\ posterior for the correlation function of a randomly selected synthetic observation to its true value.  We see that the true correlation function from the simulation (black) falls within the measurement uncertainty from the posterior distribution (purple band), as expected.

Despite this success, we are marginally able to detect a small residual bias in the recovered clustering statistics. That is to say, while the posterior distribution of all inferred summary statistics are consistent with truth, we are able to detect a bias when considering all 108 independent runs simultaneously.   We formalize this statement by considering the difference between the mean summary statistic $\langle{\vec d}\rangle$ for each of 108 posteriors and the corresponding 108 true values $\vec d_{\rm true}$.  If our posteriors are unbiased, then $\Delta \vec d$ should be consistent with zero across all 108 realizations.  We define $\vec \mu$ as the average value of $\Delta \vec d$ across all 108 simulations, and $C$ as the empirical covariance matrix of $\Delta \vec d$ from all 108 simulations.  We compute the $\chi^2$ statistic
\begin{equation}
    \chi^2 = \vec{\mu}^\top C^{-1} \vec{\mu}.
\end{equation}
Once again, we correct for the Hartlap factor during this calculation.  A large $\chi^2$ signals that despite the \Karmma\ maps being consistent with truth within noise {\it for an individual map}, when considering the full ensemble of 108 simulations we are able to detect residual biases.

\begin{figure}
    \centering
    \includegraphics[width=\columnwidth]{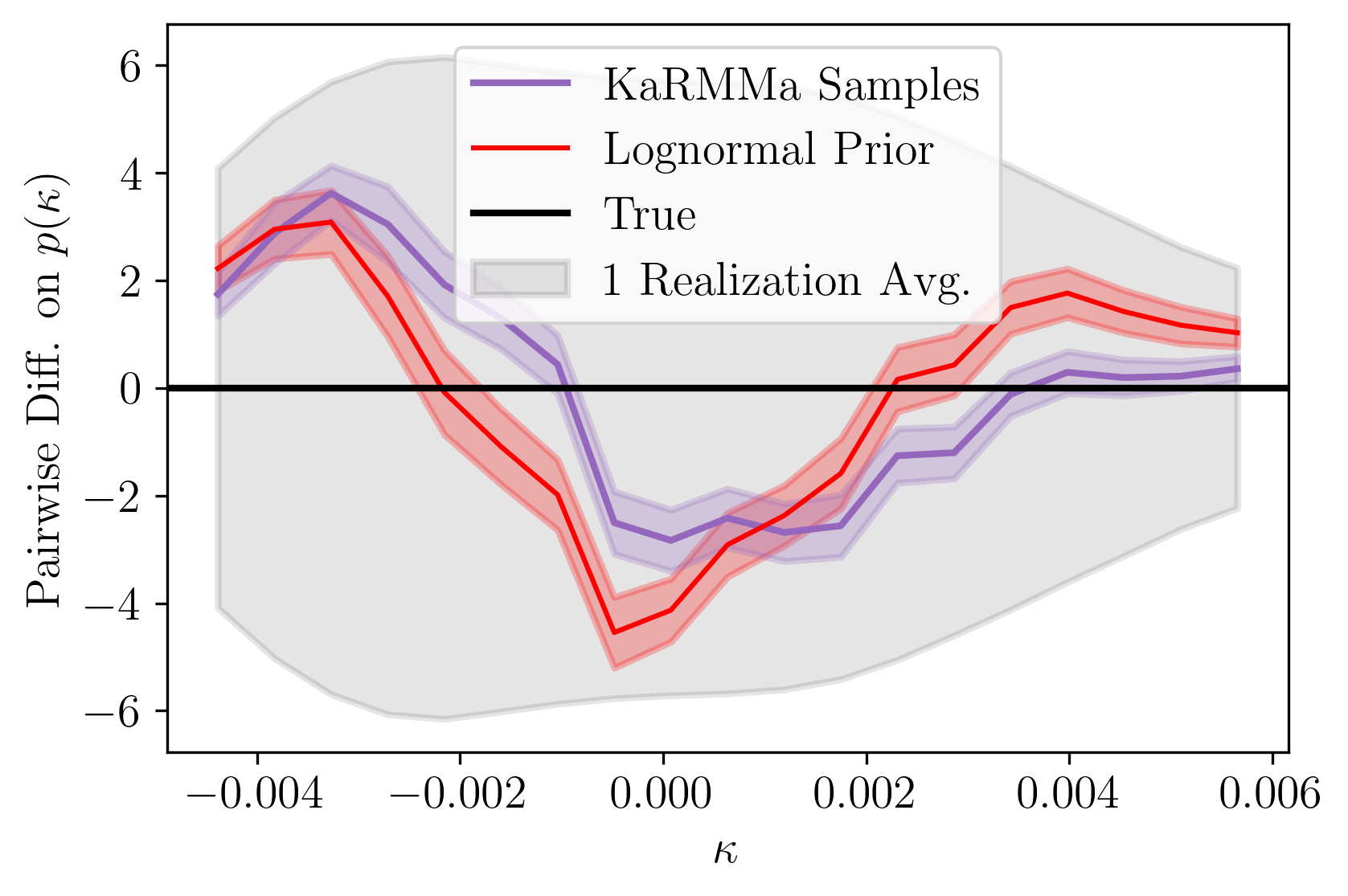}
    \caption{Difference between the mean one-point function from the \Karmma\ posterior and the true one-point function (purple line), averaged across all 108 synthetic data sets. The red line shows the average difference between the one-point function of the lognormal model used in our prior and the true one-point function.  The non-zero difference demonstrates the log-normal model provides only an approximate description of the convergence field. The purple/red bands represent the 68\% confidence interval of the mean. We detect a bias in the mean \Karmma\ one-point function at $7.2\sigma$ that is clearly driven by the lognormal model being only an approximate description of the true convergence field. Note, however, that this bias is smaller than the observational uncertainties for a single mock (grey band). A more direct comparison of the one-point functions can be seen in Figure~\ref{fig:1pt_func}.}
    \label{fig:hist_bias}
\end{figure}

As a specific example of these residual biases, the right panel of figure~\ref{fig:samp_cf_and_bias} shows the difference between the mean correlation function from the chain samples and the true correlation function, averaged across all 108 synthetic observations. The purple band is the error on the mean, estimated from the empirical covariance matrix of the differences across the 108 simulations, multiplied by 1/108 (the number of simulations). The $\chi^2/dof$ of the hypothesis that the mean difference is zero is $\chi^2/dof \approx 56/19$, and can be rejected at $\approx 2.9\sigma$. We can perform a similar calculation for the one-point function and the harmonic space two-point function. For these two cases, the $\chi^2$ values per degree of freedom are $\chi^2/dof\approx 136/19$ and $\chi^2/dof \approx 30/19$ respectively. These values correspond to a rejection of the null hypothesis at $\approx 7.2\sigma$ for the one-point function and an acceptable $\approx 1.6\sigma$ for the power spectrum. For the peak/void counts we find $\chi^2/dof$s of $\approx 260/19$ and $\approx 125/19$ corresponding to statistically significant deviations of $\approx 13.7\sigma$ and $\approx 6.6\sigma$ respectively. The large $\chi^2$ bias for the one-point function can be understood from Figure~\ref{fig:hist_bias}.  There, we show the difference between the mean one-point function from the \Karmma\ samples and the true one-point function across all 108 mock realizations. Plotted in red is the difference between the one-point function of the lognormal model used in our prior and the true one-point function. Evidently, the lognormal model fails to reproduce the true one-point function in simulations in detail. The peak/void counts show even larger biases, indicating the lognormal fails to perfectly recover higher-order statistics. In short, the use of our lognormal prior biases the recovered posteriors.  However, we emphasize that  these biases are smaller than the statistical uncertainty recovered from a single map, and are only clearly detectable when using the full simulation ensemble.

\subsection{Comparison of \Karmma\ to Kaiser--Squires}
\label{sec:comparison}

We wish to compare the \Karmma\ mass maps to those produced using the standard Kaiser--Squires algorithm.  We note that Kasier--Squires algorithm can be thought of as the maximum posterior map under a flat prior for the convergence field. To perform this comparison, we focus on the one and two-point statistics of the recovered maps, as well as the peak and void counts.  Unfortunately, the comparison is non-trivial because Kaiser--Squires is not a Bayesian reconstruction of the true mass map. Let's focus our discussion on the correlation function of the convergence field as an example.  Since the correlation function of the Kaiser--Squires convergence map is a deterministic function of the data, we could focus our investigation on the probability distribution $P(\xi|\kappa_{\rm true})$, where $\xi$ is the correlation function of the recovered mass map.  However, this distribution is insensitive to cosmic variance noise in the field, which is one of the principal sources of noise in real surveys.  We have chosen instead to compare the distributions for $\xi$ given a set of cosmological parameters $\Omega$, i.e. $P(\xi|\Omega)$. This distribution is especially easy to study since the 108 measurements produced from our simulations constitute a sampling from this distribution.  

Given that $P(\xi|\Omega)$ is the only reasonable statistic we can compute for Kaiser--Squires, we will need to construct the corresponding distribution from the \Karmma\ posteriors. Specifically, we calculate the distribution\footnote{Note that since $\xi$ is a deterministic function of the map $\kappa$, the distribution $P(\xi|\kappa)$ is a Dirac delta function.}
\begin{equation}
    P(\xi|\Omega) = \int d\gamma_\mathrm{obs} d\kappa\ P(\xi | \kappa)P(\kappa| \gamma_\mathrm{obs}, \Omega) P(\gamma_\mathrm{obs} | \Omega) 
\end{equation}
by computing $\xi$ for each sample map from the chain for each mock realization, and then combining all sample statistics into a single very large chain of sample statistics marginalized over the mock realizations. The distribution of the statistics from this combined chain consistute a sampling of the distribution $P(\xi|\Omega)$ where $\xi$ is the convergence correlation function from \Karmma\ maps. 

To help interpret the \Karmma\ posteriors for this comparison, it is worth discussing the expected distribution for $P(\xi|\Omega)$ for \Karmma\ in the limits of infinite noise and infinite signal-to-noise respectively. In the infinite signal-to-noise limit, the prior is irrelevant.  \Karmma\ will recover the underlying convergence field for each simulation exactly, and the posterior will reflect the cosmic variance in the input density fields.  That is, our posterior corresponds to the distribution of correlation functions for maps sampled from the distribution $P_\mathrm{sim}(\kappa|\Omega)$.  Conversely, in the limit of infinite noise, the posterior will reflect the $\kappa$ distribution obtained from the prior $P_0(\kappa|\Omega)$. If the log-normal prior $P_0$ were exactly identical to the simulation distribution $P_\mathrm{sim}$, then the posteriors for the infinite signal-to-noise and infinite noise limits would be identical. In other words, the distribution $P(\xi|\Omega)$ for \Karmma\ should, by construction, reflect cosmic variance uncertainties and nothing else.  In that sense, $P(\xi|\Omega)$ is not useful statistic for evaluating the efficacy of \Karmma.  However, it is the only avenue we have for comparing \Karmma\ to Kaiser--Squires on equal footing.  Of course, a similar argument can be made for the one-point distribution.

\begin{figure*}
    \centering
	\includegraphics[width=\textwidth]{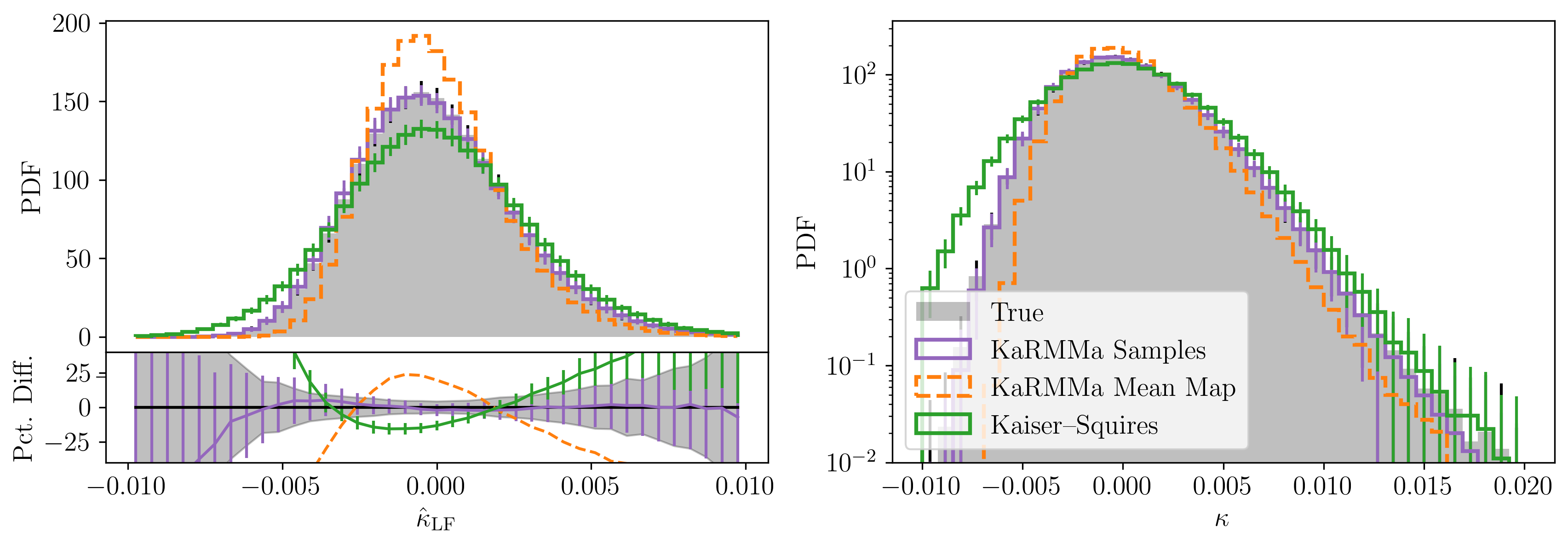}
    \caption{Top Left: The one-point function of the true convergence map (black), the Kaiser--Squires reconstruction (green), the mean \Karmma\ map (orange), and the means of the posterior estimated from the \Karmma\ samples (purple). All distributions have been averaged across the 108 simulated data sets. Bottom Left: Percent difference between the one-point distributions of each method and the true one-point distribution. Right: Comparison of one-point distributions with a log-scale. The posterior distribution from \Karmma\ results in an excellent description of the one-point function of the simulations. The error bars/bands represent the $1\sigma$ standard deviation of the computed statistics across all mock realizations.}
    \label{fig:1pt_func}
\end{figure*}

With the nuances of this comparison established, we begin by looking at the one-point distributions. Figure \ref{fig:1pt_func} shows that the mean one-point function from the Kaiser--Squires reconstruction (green) is much too broad compared with the mean one-point function from the input simulated maps (black).  The Kaiser--Squires one-point function is also much more symmetric than the clearly non-Gaussian true one-point function \citep[see e.g. fig. 8 in][]{y3_mass_map}. By contrast, we see that the mean one-point function of \Karmma\ sample maps (purple) is nearly identical to that of the true map, as expected based on our discussion is section~\ref{sec:posterior_validation}.  If we consider the distribution of the mean \Karmma\ maps instead (orange), we find that the resulting one-point function is non-Gaussian but much too narrow, reflecting the ``missing variance'' due to unresolved modes being zeroed-out in the averaging procedure. 

\begin{figure}
    \centering
	\includegraphics[width=\columnwidth]{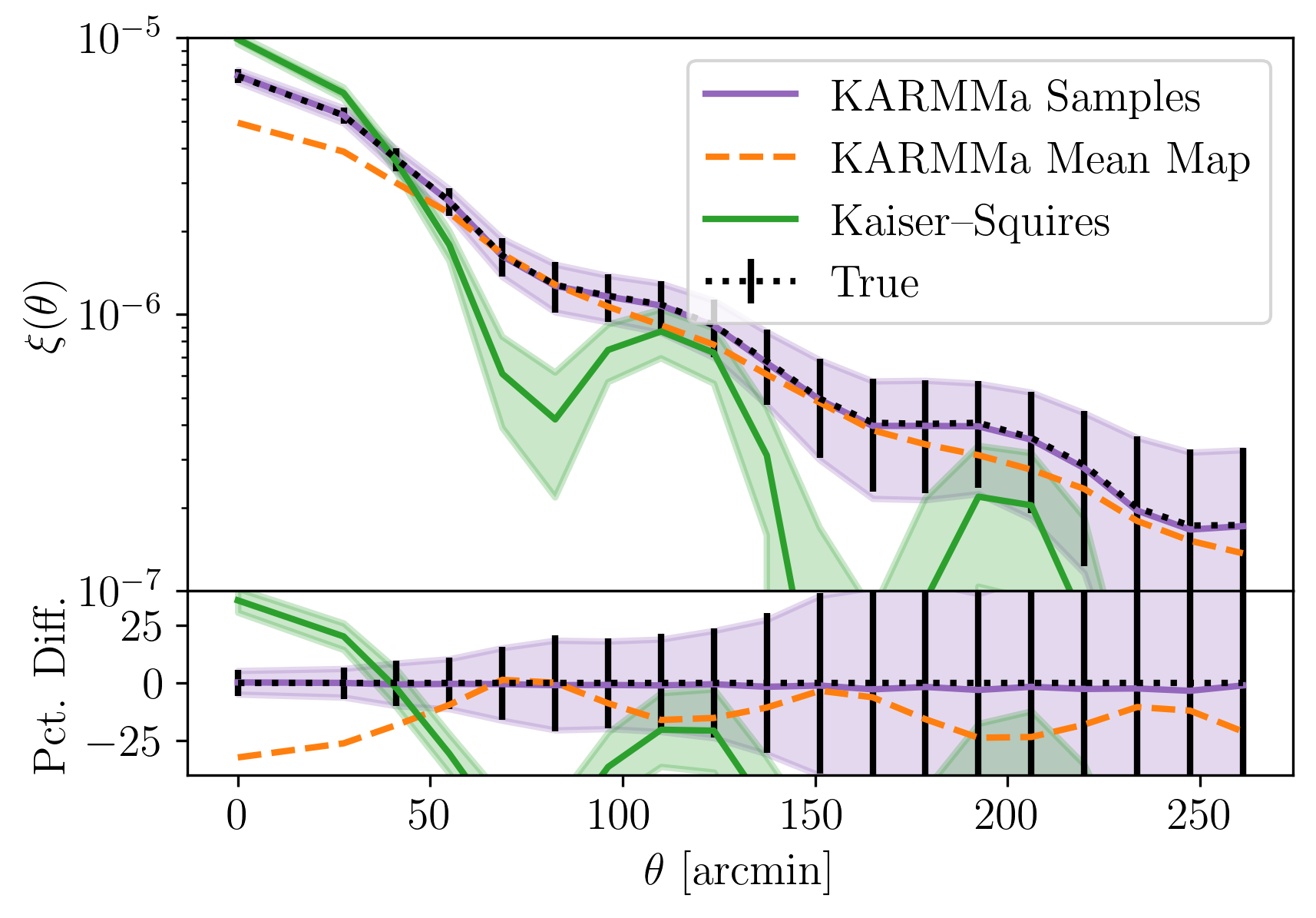}
    \caption{Top: Two-point angular correlation functions of each reconstruction method compared to the true correlation function in the simulations. Bottom: Percent difference between the correlation function and the true two-point function. The \Karmma\ samples correctly recover the true correlation function in the simulation, whereas the two-point function of the Kaiser--Squires reconstruction is grossly biased.}
    \label{fig:2pt_func}
\end{figure}

\begin{figure}
    \centering
    \includegraphics[width=\columnwidth]{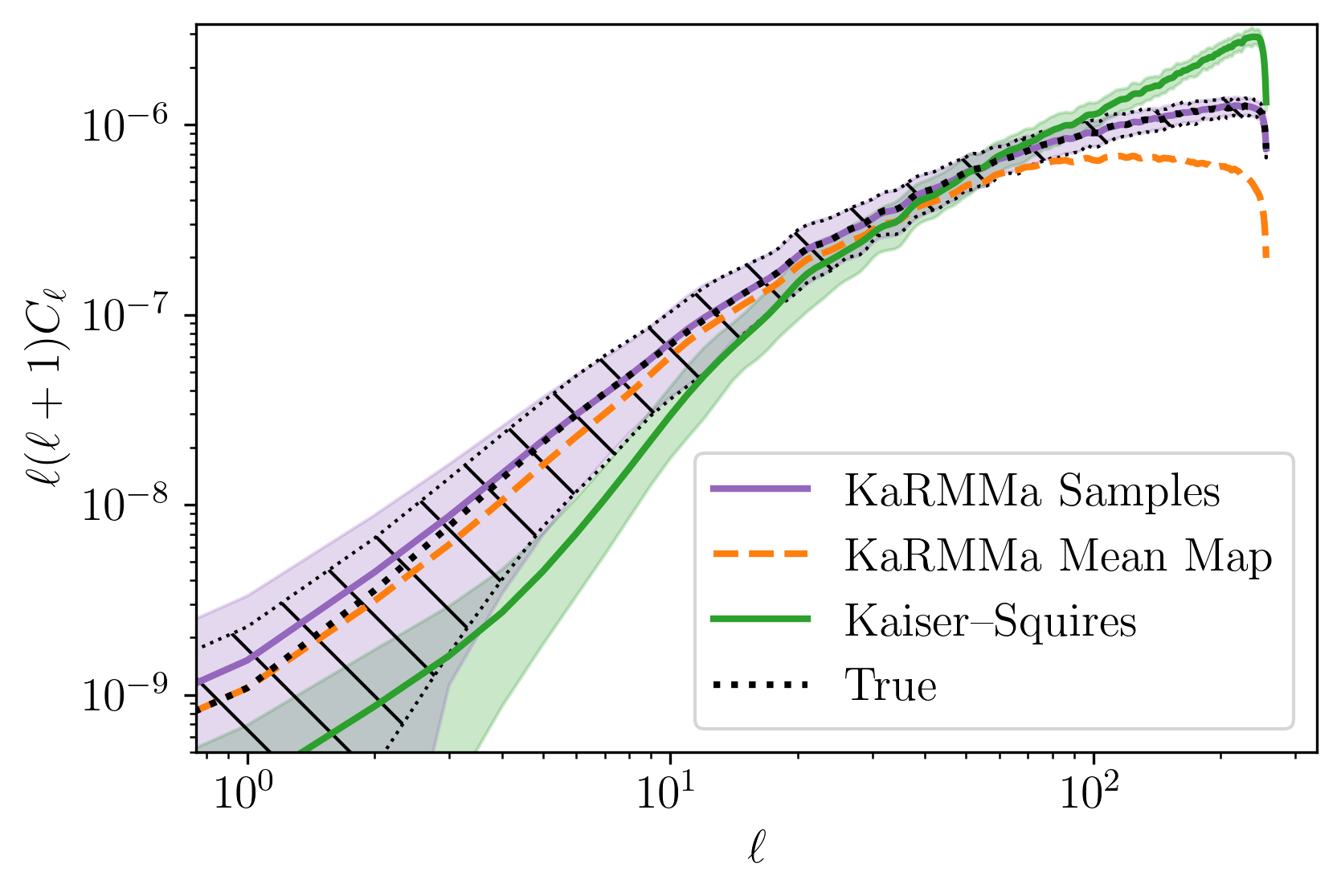}
    \caption{Power spectra from each reconstruction method compared to the true power spectrum from simulations. As with the correlation function, we find that the \Karmma\ samples easily outperform the Kaiser--Squires reconstruction, resulting in very nearly unbiased power-spectrum estimates.  This figure also highlights the suppression of power at small scales in the mean \Karmma\ map, as well as the artificially high small scale structure in the Kaiser--Squires reconstruction. Interestingly, Kaiser--Squires also fails to recover the correct amount of large scale power.}
    \label{fig:cl_plot}
\end{figure}

Figures~\ref{fig:2pt_func} and~\ref{fig:cl_plot} show the correlation function and power spectrum for each of our four types of maps: true, Kaiser--Squires, mean \Karmma\ map, and \Karmma\ posterior.  We find that the correlation function of the Kaiser--Squires reconstruction (green) is biased, consistent with our observations that the spatial structure of these maps ``looks wrong'' by eye.  In particular, we note the Kaiser--Squires reconstruction exhibits excess noise at small scales, and under-estimates the power at large scales due to leakage into shear B-modes \citep[for further discussion, see][]{y3_mass_map}. The mean \Karmma\ map (orange) is a clear improvement over Kaiser--Squires, but the smoothing due to the averaging of the maps biases reduces the correlation function and power-spectrum at small scales.  This makes sense, since it is precisely the small-scale modes that are unresolved. Finally, the mean of the \Karmma\ posterior for the correlation function (purple) is a near-perfect match to the true correlation function (black).

\begin{figure*}
    \centering
    \includegraphics[width=\textwidth]{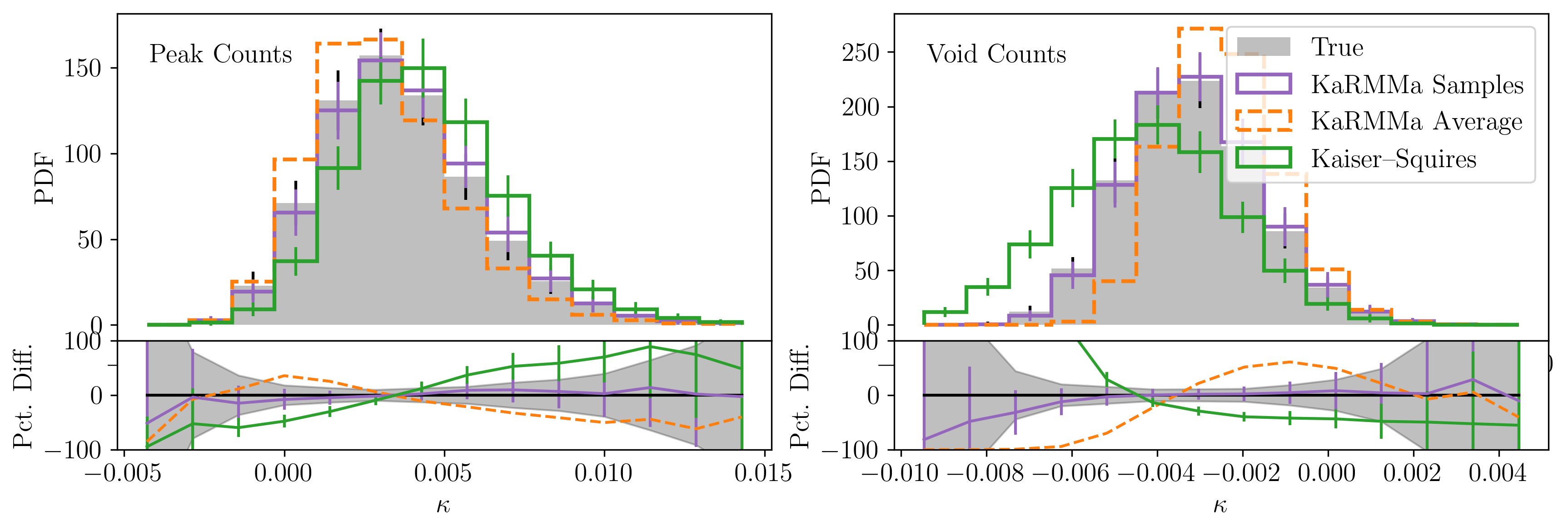}
    \caption{{\it Left:} Peak counts from each reconstruction method compared to the true peak counts from simulations, as labeled. {\it Right:} Similar to the left panel, but for void counts. The distribution of peaks and voids in the Kaiser--Squires map is heavily skewed compared to truth due to the high frequency noise in these maps.  By contrast, the \karmma\ maps have peak and void counts that are within noise of the the true distributions.}
    \label{fig:pc_plot}
\end{figure*}

Peak/void counts are widely used in cosmology as they are simple to compute and capture some of the non-Gaussian information contained within the field. In this paper, we define peaks/voids as pixels in the map whose $\kappa$ value is greater/less than the value of $\kappa$ in all immediately neighboring pixels. In figure \ref{fig:pc_plot}, we compare the peak/void distributions from each reconstruction method to that of the true distributions from simulations. We see that the \karmma\ posterior for the peak and void counts match the true counts in the simulations within noise.  By contrast, the Kaiser--Squires reconstruction has peak/void counts which are skewed towards higher/lower values of $\kappa$ than the true distribution. We attribute this excess to the excess high frequency noise in the $\kappa$ distribution in the Kaiser--Squires maps.

\subsection{Comparison of \Karmma\ to Gaussian Prior}

The aim of \Karmma\ is to provide a fast tool for reconstructing mass maps that also accurately recovers the correct map statistics. Key to achieving this goal is the choice of the lognormal model. The lognormal model provides a simple yet relatively accurate approximation of the non-Gaussianities of the convergence field. By contrast, while a Gaussian prior should recover the correct two-point statistics of the map, it should fail to recover non-Gaussian statistics.  In this section we compare the performance of \Karmma\ when using lognormal vs gaussian priors.   We are particularly interested in how well each method reconstructs non-Gaussian statistics.

\begin{figure*}
    \centering
    \includegraphics[width=\textwidth]{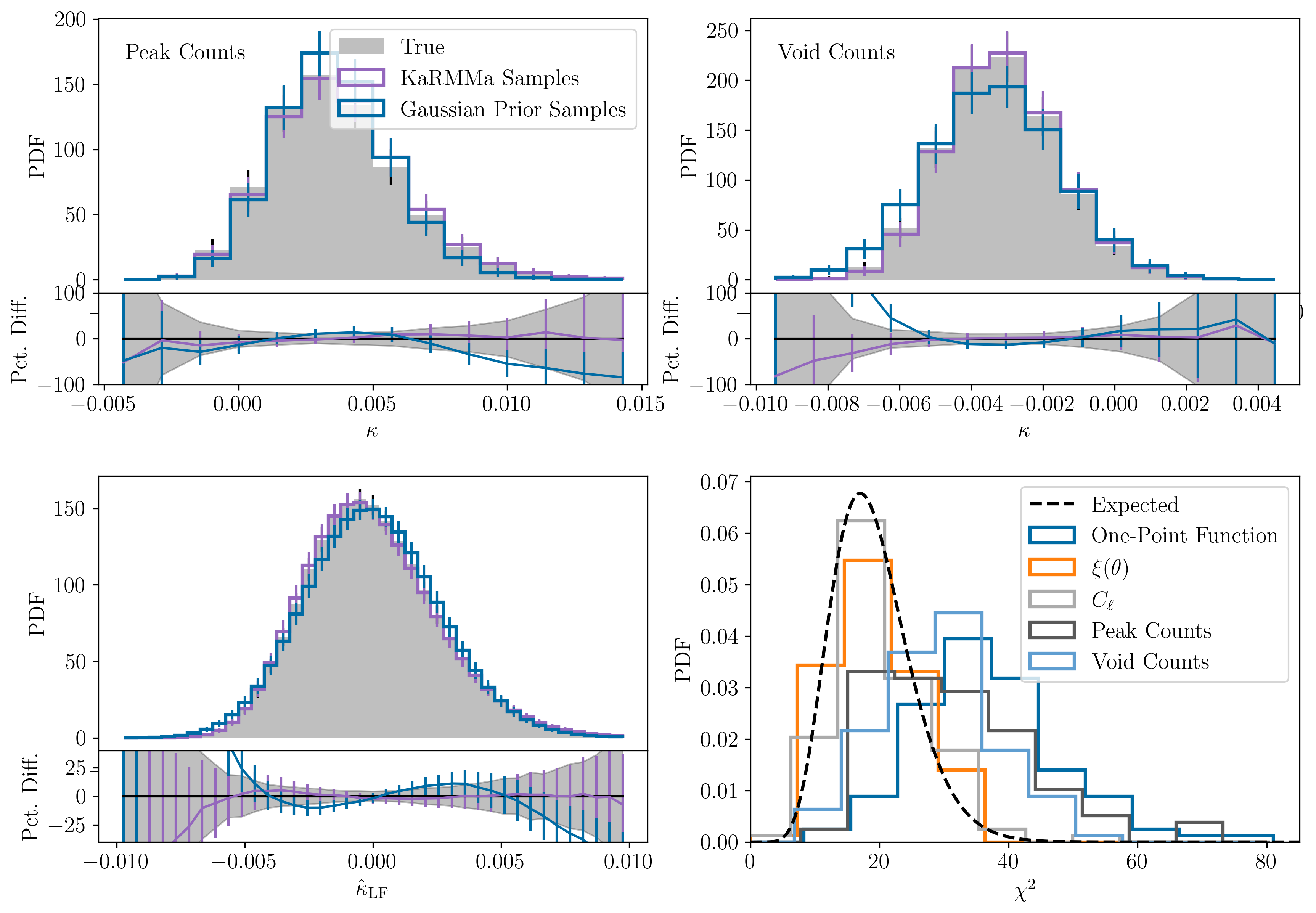}
    \caption{Comparison of non-Gaussian statistics from the \Karmma\ posterior samples to samples obtained using a Gaussian prior. Top Left/Right: Peak counts/void counts. Bottom Left: One point function. Bottom Right: Chi-squared distributions for the various Gaussian and non-Gaussian summary statistics for the samples generated using a Gaussian prior. As expected, the two-point functions of the Gaussian maps are consistent with the a chi-squared distribution. However, the one-point function and peak/void counts are not properly reconstructed when using a Gaussian prior, resulting in significantly biased chi-squared distributions.}
    \label{fig:gauss_prior}
\end{figure*}

Figure \ref{fig:gauss_prior} compares the peak/void counts and one-point functions from the \Karmma\ samples (purple) and Gaussian prior samples (blue) to the true distribution (black) from simulations. From these plots, it is obvious that the use of a Gaussian prior significantly biases the resulting maps.  Moreover, this bias is highly significant, as evidenced by the resulting $\chi^2$ distributions of the inferred statistics compared to truth.  By comparison, the chi-squared distributions from the \Karmma\ samples (see fig. \ref{fig:chisqs}) are virtually unbiased.  As expected, the two-point statistics from the Gaussian prior samples do follow the expected chi-squared distribution. That is, as is the case for the lognormal model, the adoption of a Gaussian prior does not bias the inferred two point functions.

\section{Discussion}
\label{sec:discussion}

As mentioned in section \ref{sec:intro}, many other methods have been developed to perform Bayesian mass map reconstruction. Notably, \cite{B_hm_2017} also used a 3D lognormal prior to forward model the shear field. The authors demonstrated that the lognormal prior performed better than the Gaussian prior, particularly at the small non-linear scales. Furthermore, their approach allowed the authors to incorporate redshift uncertainties for individual sources rather than requiring a pixelization scheme on the sky or tomographic binning. However, \cite{klypin_2018} showed that the lognormal model is not an accurate description of the 3D density field and fails at small, very non-linear scales. At these scales, more accurate models may be required. For instance, \cite{Porqueres_2021} aims to reconstruct the lensing signal by forward modelling the non-linearly evolved density field from simulations. Evidently, this simulation-based approach can more accurately model the physics of the non-linear scales than the simple lognormal model.

The approaches from \cite{B_hm_2017} and \cite{Porqueres_2021} are computationally expensive as they involve reconstructing the full 3D density field in order to recover the lensing signal. In particular, \cite{B_hm_2017} only performed maximum a posteriori (MAP) reconstruction by maximizing the posterior. As highlighted in this work, the MAP reconstruction suffers from suppressed power in prior-dominated regimes and therefore results in an unphysical density field with incorrect statistical properties. By contrast, as shown in section \ref{sec:comparison}, sampling the posterior produces maps with accurate statistical properties. An additional consequence of reconstructing the MAP estimate is that uncertainties and covariances are not trivial to recover. The authors solved this by using the Laplace approximation (which assumes that the posterior is Gaussian around the MAP point) to estimate the diagonal of the covariance matrix of the map.

In our approach, we aimed to develop a fast method for recovering the weak lensing signal by directly reconstructing the 2D mass maps rather than reconstructing the entire 3D density field. Similar 2D efforts have appeared in the literature, for instance Wiener filtered maps \citep{Jeffrey_2018}.  Here again, a key advantage of our approach is the fact that we sample the full posterior distribution of the maps.  Consequently, clustering statistics can be estimated directly from the maps in a nearly unbiased way, while uncertainties in the same are trivial to compute.  In particular, section \ref{sec:posterior_validation} demonstrates that the uncertainties and covariances recovered from our method are indeed accurate.

Another noteworthy 2D Bayesian reconstruction technique is GLIMPSE \citep{Leonard_2014, Jeffrey_2018}. GLIMPSE performs a MAP reconstruction and imposes a prior of sparsity on the wavelet coefficients of the convergence map. The wavelet filter in the GLIMPSE algorithm is chosen with the goal of adequately modelling halos and small scale features. However, the GLIMPSE reconstruction ultimately results in mass maps with incorrect clustering statistics due to the reliance on MAP estimates and the fact that the sparsity prior, while it serves to regularize the forward model, does not ensure that small scales statistics are correctly recovered. By comparison, the prior in our reconstruction technique was selected to provide a simple yet relatively accurate description of the non-Gaussian convergence field.
Recently, machine learning methods have been shown to hold great promise in their ability to generate realistic lensing fields \citep[e.g.][]{remy2020probabilistic}. These are typically limited to small flat sky patches, though this is rapidly changing \citep[][]{deepsphere, deepspheregan}.

It is our view that this broad range of methods is to the benefit of the weak lensing program of the community as a whole.  There are clear advantages and disadvantages to each approach, and it seems likely that ultimately ideas drawn from several of these algorithm will be adopted as standard by the community in the future.


\section{Summary and Conclusions}
\label{sec:conclusions}

In this work we present \Karmma, a new method for reconstructing mass maps from weak lensing shear observations. \Karmma\ provides a fully Bayesian reconstruction with sample maps drawn from the posterior relying on a physically motivated lognormal prior for the convergence field $\kappa$.  \Karmma\ produces a full library of sample maps from the posterior distribution, making it ideal for use in cross-correlation studies with other astrophysical maps \citep[for example, see][]{Chang_2018}. Because the \Karmma\ samples accurately reflect the measurement uncertainty (see section~\ref{sec:posterior_validation}), estimation of observational uncertainties in cross-correlation studies becomes trivial: one need only cross-correlate the data of interest with the \Karmma\ library of posterior samples. Further, we demonstrated that the two-point functions of the posterior maps are very nearly unbiased relative to truth.  Indeed, any residual biases are significantly smaller than statistical uncertainties (see right panel in Figure~\ref{fig:samp_cf_and_bias}).  

In section~\ref{sec:comparison} we compare the \Karmma\ and Kaiser--Squires mass map reconstructions on a suite of dark matter simulations. We demonstrate that the best point estimate from \Karmma\ outperforms the Kaiser--Squires reconstruction in that it exhibits a narrower distribution of residuals when compared to the true mass map. Moreover, unlike the Kaiser--Squires reconstruction, the \Karmma\ posteriors recover nearly unbiased clustering statistics.

There may be use cases in which the use of the mean \Karmma\ map is more desirable than the use of the full posterior, for instance, if a single ``best-map'' estimate is desired. In this context, the Kaiser--Squires reconstruction could also be a suitable choice, though we believe that for most if not all applications the mean \Karmma\ map is preferable to the Kaiser--Squires reconstruction.  An obvious possible exception to this rule is cosmological investigations (see below).

While this iteration of \Karmma\ has successfully demonstrated the value of our forward modeling approach, our current algorithm is limited in important ways. First, \Karmma\ requires a cosmology to be specified for the prior, thus preventing our mass maps from being used for constraining cosmology without significant additional work.  However, one could imagine calibrating the incurred bias as a function of cosmology through a response analysis \citep{Seljak_2017, horowitz2019}. Second, \Karmma\ can only perform reconstruction for one tomographic bin at a time, and therefore cannot perform a joint reconstruction across multiple tomographic bins. Third, the current parameterization of \Karmma\ restricts our reconstruction to relatively low resolutions. Future work on this method will address all three issues, allowing \Karmma\ to perform multi-bin mass map reconstruction at arcminute resolution while simultaneously sampling mass maps and cosmology.

The current version of the \Karmma\ code is available on GitHub at \url{https://github.com/pierfied/karmma}. This code is provided as a Python package and utilizes PyTorch to drastically improve sampling speed by taking advantage of CUDA GPUs.


\section*{Acknowledgements}

PF and ER are supported by NSF grant 2009401.  ER is further supported by DOE grant DE-SC0009913, and by a Cottrell Scholar award.

\section*{Data Availability}

The data underlying this article was derived from \cite{Takahashi_2017} which can be accessed at: \url{http://cosmo.phys.hirosaki-u.ac.jp/takahasi/allskyraytracing/}. The derived data generated in this research will be shared on reasonable request to the corresponding author.



\bibliographystyle{mnras}
\bibliography{references} 



\appendix

\section{Calibrating Numerical Systematics}
\label{sec:bias}

As discussed in \ref{sec:res_err}, \Karmma\ HEALPix maps are restricted to well-resolved $\ell$ modes, that is $\ell \le 2\ n_\mathrm{side}$. As a result, power from modes with $\ell > 2\ n_\mathrm{side}$ are lost in the forward-modeled shear field. However, as the true shears include these modes, simply ignoring them would bias the sample maps. Additionally, because the convergence field is not reconstructed over the full-sky, the predicted shear field also includes masking effects that must be accounted for (see section \ref{sec:mask_effects}). We address these two systematics by including them as noise terms in our reconstruction.

These numerical systematics can be expressed as linear operations that must be tacked on to our model.   That is, the true shear field can be written as $\vec{\gamma}_t = \hat{\gamma} + \vec{\epsilon}$ where $\hat{\gamma}$ is the forward-modelled shear field and $\vec{\epsilon}$ is the noise term due to the above systematics. In theory, this noise term is inherently non-Gaussian due to the non-Gaussianity of the mass maps. However, for simplicity we will assume this additional term is Gaussian.  Empirically, we find that this approximation is sufficient.

In our reconstruction, we model the systematics term $\vec \epsilon$ as a zero-centered Gaussian and calibrate its covariance matrix using 10,000 lognormally generated mocks with the same cosmology as the simulations used in section \ref{sec:sim_tests}. For each mock, we apply the $\ell_\mathrm{max}$ cut and mask to the true $\kappa$ maps, then use the Kaiser--Squires transformation to get the predicted shear field including systematics $\hat{\gamma}$. We then compare these predictions to the true shear field to get 10,000 samples of the systematics vector, $\vec{\epsilon} = \vec{\gamma}_t - \hat{\gamma}$. These samples are then used to estimate the covariance matrix for the systematics. Finally the numerical systematics are accounted for during inference by adding the systematics covariance matrix to the diagonal shape-noise covariance matrix in the likelihood term.

The DES Y1 footprint as used in our tests contains approximately 8,000 pixels at the resolution of $n_\mathrm{side}=128$. This would require on the order of $10^8$ parameters for the systematics covariance. However, the empirical covariance matrix was estimated using only 10,000 samples, which would result in an unstable, numerically noise covariance matrix. An attempt to use shrinkage \citep{Chen_2010} to regularize the matrix proved unsuccessful: a small residual bias in the recovered correlation functions was apparent in this case.  Instead, we found that we could produce a better approximation of the systematics covariance matrix by taking advantage of the expected structure of the matrix. The noise introduced by masking effects should be most strongly correlated with nearby pixels. Additionally, the unmodelled modes with $\ell > \ell_\mathrm{max}$ are predominantly at the pixel/subpixel scale and, should therefore not be strongly correlated with distant pixels. Consequently, we modify our estimated covariance matrix by setting the covariance for all pairs of pixels separated by more than 250 arcmin to zero. We found this to be the most effective treatment of the above theoretical systematics.  


\bsp	
\label{lastpage}
\end{document}